\documentclass[final,5p,times,twocolumn]{elsarticle}

\usepackage{graphicx}
\usepackage{amssymb}

\newcommand{\ICC}[0]{ICC$^\star$}

\renewcommand{\vec}[1]{\mathbf{ #1}}

\renewcommand{\d}{\mathrm d}
\def\es{{\sf ESPResSo}}

\journal{Computer Physics Communications}

\begin{document}

\begin{frontmatter}



\title{The \ICC\ Algorithm: A fast way to include dielectric boundary effects
into molecular dynamics simulations}


\author[S]{Stefan Kesselheim}
\address[S]{Institute of Computational Physics, University of Stuttgart, Pfaffenwaldring 27, D-70569 Stuttgart, Germany}
\author[T]{Marcello Sega}
\address[T]{Department of Physics and INFN, University of Trento, via Sommarive 14, I-38123 Trento, Italy}
\author[S]{Christian Holm}
\ead{holm@icp.uni-stuttgart.de}

\begin{abstract}
  We employ a fast and accurate algorithm to treat dielectric
  interfaces within molecular dynamics simulations and demonstrate the
  importance of dielectric boundary forces (DBFs) in two systems of
  interests in soft-condensed matter science. We investigate a salt
  solution confined to a slit pore, and a model of a DNA fragment
  translocating thorugh a narrow pore.
\end{abstract}

\end{frontmatter}


\section{Introduction}
Coarse-grained models are widely employed in computer simulations of
soft-condensed matter systems because of the significant computational
speed-up granted by the reduction in the number of degrees of
freedom. A common coarse-graining approach is to employ an implicit
solvent approach, thereby removing completely all solvent molecules.
In the simplest scheme the polarizability of an aqueous solvent is
taken into account by setting the dielectric constant to 80, thus
reducing the electrostatic interaction by the same factor.  This
approach, however, fails in presence of interfaces between the solvent
and materials of significantly different polarizability.  Various
methods to solve the Poisson equation with inhomogenous dielectric
permittivity are available, and have been employed in different
contexts, such as in
Refs.~\cite{warwicker82a,levitt78a,Honig93,miertus81a,shaw85a,Chipman04,boda04b,allen01a,tomasi94a,Bardhan09a,Lu06}.
One of the possibilities consists in solving the boundary integral
equations using boundary element methods. This approach is
particularly useful when the boundary integral problem is formulated
in terms of an system of implicit linear equations which express the
induced surface polarization charges in terms of the electric
field\cite{Lu06,Fenley96,Zauhar96,Purisima98,Bordner03,bharadwaj95a}.
With the \ICC\ algorithm\cite{tyagi10a} (Induced Charge Calculation
with fast Coulomb Solvers) we enhanced the capabilities of different,
widely employed electrostatic solvers with an iterative scheme to
solve this set of equations. Noticeably, the procedure automatically
yields solutions which obey the boundary conditions of the underlying
Coulomb solvers, which is important to reduce the necessary system
size.

In this article we review the algorithm and present two applications in the context of
soft-condensed matter. The first one is the test case scenario of an electrolyte confined
between two walls. In the second application we calculate the free energy barrier which a
model DNA fragment has to overcome in order to be transported through a synthetic nanopore,
for two different ionic strength conditions.

\section{The \ICC\ Algorithm}
The goal of the \ICC algorithm is to solve the Poisson equation for an inhomogenous
dielectric. The Poisson equation in cgs units reads as:
  \begin{equation}
  \nabla \cdot \left(\varepsilon \nabla \Phi\right) = - 4 \pi \rho_\mathrm{ext},
  \label{eq:poisson}
\end{equation} 
Let us suppose for simplicity that there is only one interface between
the two regions 1 and 2
of different dielectric permittivities $\varepsilon_1$ and $\varepsilon_2$.
In order to express the boundary conditions in terms of induced surface charges,
we integrate eq.~\ref{eq:poisson} over a pillbox small enough that the inhomogenity 
of the electric field through its caps is negligible. Then the induced charge in the pillbox can be expressed as a surface
integral
\begin{equation}
  \begin{array}{ccl}
  \displaystyle  { 
  q_\mathrm{ind} } & \displaystyle { = } & 
  \displaystyle  { 
    \frac{1}{4 \pi} \displaystyle  \oint \d \vec{A} \cdot  \varepsilon \nabla \Phi } \\ 
   &\displaystyle { = } & 
  \displaystyle  { 
    \frac{A}{4 \pi} \left( \varepsilon_1 \vec{E}_1 \cdot \vec{n} - \varepsilon_2 \vec{E}_2 \cdot \vec{n} \right),
    \label{eq:qind}}
  \end{array} 
\end{equation}
where $\vec{n}$ is the surface normal, conventionally pointing from
region 1 into region 2, and $\vec{E_{1/2}}$ is the electric field in
medium 1 or 2 in close proximity of the interface.  The last equation
is valid in the limit of a surface element of infinitesimal area $A$.

The field $\vec{E_{1/2}}$ at a given position on the surface can be written as a superposition of the field generated by
the charged surface element in the very same location and of the field generated by other external and induced charges in the
system. We denote the second contribution
by $\vec{E}$ and the charge density on the surface element by $\sigma=q_\mathrm{ind}/A$.
\begin{equation}
  \vec{E_{1/2}} = \vec{E} \pm 2 \pi/\varepsilon_1 \sigma  \vec{n}.\label{eq:Esurf}
\end{equation} 
By combining Eqns.(\ref{eq:qind}) and (\ref{eq:Esurf}) one obtains an expression for the
induced surface charge:
\begin{eqnarray}
  \sigma = \frac{\varepsilon_1}{2 \pi} \frac{\varepsilon_1 - \varepsilon_2}{\varepsilon_1 + \varepsilon_2} \vec{E}\cdot\vec{n}.\label{eq:bie}
\end{eqnarray}
This boundary integral formulation can be implemented in a numerical boundary
element scheme by assigning charge elements on a grid which discretizes the
interface. Eq.(\ref{eq:bie}) then is a set of coupled equations, which can
be solved with different approaches. In the \ICC\ implementation we decided to use
a successive over-relaxation (SOR) scheme to solve them iteratively. When all
boundary elements are approximated as point charges 
it is possible to determine $\vec{E}$ with widely used fast Coulomb solvers that take into account
the desired periodic boundary conditions of the system, and simultaneously determine
the surface charge density to arbitrary precision. In every step of the iterative
scheme the new guess for the charge at each surface element $q^i_\mathrm{new}$ is
determined from the previous value with the following equation:
\begin{eqnarray}
   q^{i}_\mathrm{new} = \left(1-\lambda\right) q^{i}_\mathrm{old} + \lambda A^i \sigma^i,
\end{eqnarray}
where $\sigma_i$ is obtained from Eq.~(\ref{eq:bie}). 
Here $\lambda$ is a free parameter in the range between 0 and 2. With a value of
$\lambda = 0.9$ we obtained a satisfactory speed of convergence and no stability
issues in all performed simulations. Although more refined approximations than the
use of point charges are available\cite{Bardhan09a}, our choice allowed us to
readily implement the \ICC\ algorithm in the \es\ molecular dynamics package.  In
\es\  it is currently possible to use \ICC\ with the following electrostatic solvers:
 P3M\cite{hockney81a,deserno98a,deserno98b}, ELC\cite{arnold02c,arnold02d},
MMM2D\cite{arnold02a}, and MMM1D\cite{arnold05b,brodka06a}.

Finally it should be noted that in a typical coarse-grained simulation, the
small change of particle positions in one simulation timestep leads only to a small
change in the boundary element charge, so that each \ICC\ update usually needs
only 1-3 steps of the iterative algorithm.

\section{Electrolyte in a Slit Geometry}
As a first example of the use of \ICC\ we report here the results of a coarse-grained
Langevin dynamics simulation of a salt solution confined to a slit pore. We employ
the \emph{restricted primitive model} for electrolytes, namely representing ions as repulsive charged spheres
embedded in a dielectric continuum with $\varepsilon = 80$, and use the \ICC\ algorithm
to investigate the role of dielectric contrast between the solution and the wall
material.  The temperature $T$ was set to \mbox{300 K}, and the Bjerrum length to
\mbox{0.71 nm}, corresponding to that of water at room temperature. An excluded
volume interaction between ions was set up by means of a Weeks-Chandler-Anderson
potential
\begin{equation}
  U_\mathrm{LJ}(r)= 
  \left\{
  \begin{array}{rl}4 \epsilon_{\mathrm{LJ}} \left[\left(\frac{\sigma}{r}\right)^{12} - 
      \left(\frac{\sigma}{r}\right)^{6}\right] + \epsilon_\mathrm{LJ}   & \mathrm{if }r < 2^{1/6}\sigma \\
0 & \mathrm{otherwise,}
\end{array}
\right.
\end{equation}
with interaction strength $ \epsilon_{\mathrm{LJ }}$ equal to 1$k_BT$ ($k_B$ being Boltzmann`s
constant) and a cutoff radius $r_c=2^{1/6}\sigma$, where $\sigma=0.284$~nm.
The slit pore has a dimension of $7\times 7\times 3.5$ nm, with periodic boundary
conditions applied along the two longer edges, and has been filled by 50 positive
and 50 negative ions, reaching an approximate concentration of 200 mmol/l. In
order to investigate the effect of the dielectric mismatch, the dielectric
permittivity on the left wall of the pore has been set to 6400, while no dielectric
contrast has been set for the solvent/right wall interface.
\begin{figure}[t!]
  \includegraphics[width=\columnwidth]{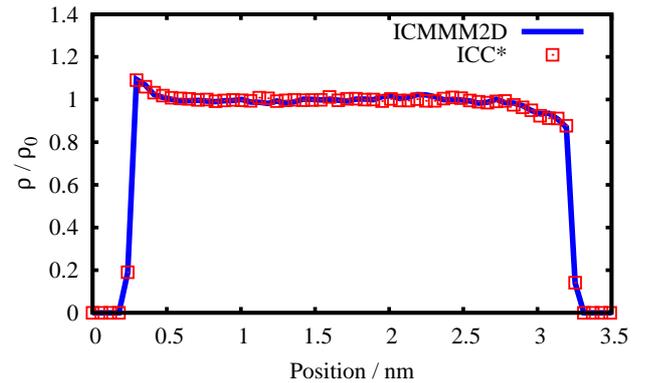}
  \caption{\label{fig:salt} 
  Density profile of ions in a slit geometry. The dielectric constant of the left,
  middle, and right regions are 6400, 80 and 80, respectively. The ICMMM2D (solid
  line) and \ICC\ results (squares) are reported.}
\end{figure}
In Fig.\ref{fig:salt} we present a comparison between the density profile of ions
in the slit, obtained with \ICC\ in combination with MMM2D as an electrostatic
solver, and with ICMMM2D. The latter method is an extension of the MMM2D algorithm,
that takes into account the presence of dielectric mismatch for flat boundaries
by means of image charges. As it is apparent from the results in Fig.\ref{fig:salt},
the use of a discretization grid of $10\times 10\times$ elements for \ICC\ is
enough to obtain results compatible with the ones obtained by using ICMMM2D within a relative error
of 1\%. The target precision of MMM2D in both cases has been set to 0.1\%. A
depletion layer due to entropic reasons is observed at the right interface, where
no dielectric contrast is present. The opposite phenomenon is observed at the left
interface, where the presence of induced surface charges introduces an effective
attractive force which compensates the entropic depletion layer and generates a
local density increase.

\section{Potential of Mean Force of a DNA segment in a Nanopore}
Experiments on DNA translocation through biological and synthetic pores have
recently attracted a lot of attention. Detailed descriptions of experiments
are available e.g.~in refs.~\cite{meller01a, dekker07a, howorka09a, storm05X}.
It has been shown that the translocation rate of DNA through a pore depends strongly
on the ionic strength of the buffer, hence indicating an electrostatic contribution
to the translocation free energy barrier. Since a highly charged object like
DNA is repelled from walls that are less polarizable than water, we 
investigate to which extent the DBFs influence the translocation free energy barrier.

We modelled a synthetic nanopore using hard walls with a dielectric constant
of 2, choosing the pore diameter and length to be 5 and 8 nm, respectively, as
they fall within the range of experimental samples.  On this length scale double
stranded DNA (dsDNA) is stiff, as it has a persistence length of $\simeq 50$ nm
at physiological conditions. Therefore it is justified to model it as a series
of beads (500, in this work) constrained on the pore axis. The inter-bead distance
is kept fixed, and each of the beads bears a charge so that the linear charge
density of 0.17$e$/nm, characteristic of dsDNA, is reproduced. 

Ions and counterions are represented explicitly. They interact mutually and with
the DNA beads through a WCA potential having $\sigma=0.425$ and $\sigma=1.215$ nm,
respectively, corresponding to a DNA diameter of 2 nm. 
The WCA interaction strength is set for both cases to 1 $k_BT$. The
electrostatic interactions are calculated in full 3D periodicity with the P3M
algorithm. We employ a cubic simulation box with a 20 nm long edge. To mimic the
case of low salt concentration,
 only monovalent counterions were introduced in the box so that the DNA molecule is neutralized.
In an additional simulation, a finite salt concentration of 100 mmol/l was added.  All
simulations were performed both with and without application of the \ICC\ algorithm to
investigate the influence of DBFs.
\begin{figure}[t!]
\includegraphics[width=\columnwidth]{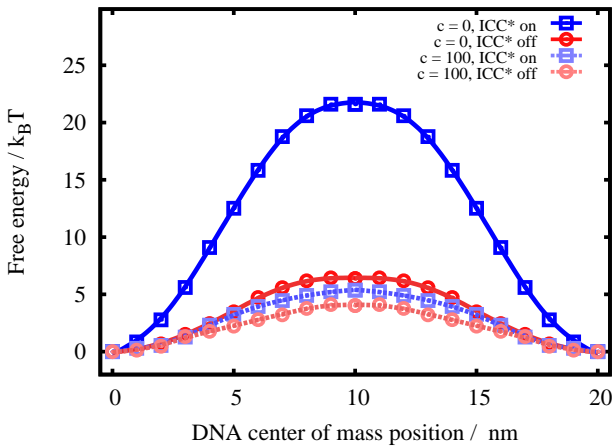}
\caption{\label{fig:pomf}  
Free energy profiles for a DNA segment in the salt free case (solid lines, $c=0$ mmol/l) and with 
a salt concentration $c=100$ mmol/l (dashed lines). Both cases are reported with (squares) and without
(circles) taking into account polarization effects with \ICC. Statistical error bars are
smaller than the symbols.
}
\end{figure}

It is straightforward to calculate then the free energy barrier
by computing the potential of mean force (PMF) acting on the center of mass of the
model DNA along the pore axis. For this reaction coordinate the Fixman
potential\cite{fixman78a} is constant, and the PMF can be obtained by numerical
integration of the mean force.
The obtained PMFs are shown in Fig.~\ref{fig:pomf}. The free energy barrier in the
salt-free case is strikingly higher (increasing to approximately $20 k_BT$) when DBFs are taking into account using \ICC, in
comparison to the case when DBFs are not considered.
On the contrary, at a salt concentration of 100 mmol/l, the barrier
increase is less pronounced, and the curves obtained with or without taking into
account DBFs show a comparable pattern with a barrier height of about $4 k_BT$.
The presence of a barrier for the higher salt concentration case, as well as in
the case when no DBFs are considered, can be explained by the steric confinement
of the counterion cloud in the nanopore. On the contrary, at low salt concentration
the Coulomb interaction is not screened, and the effect of DBFs is maximized, leading
to the observed higher free energy barrier.

\section{Conclusion}
We have presented the \ICC\ algorithm and have shown that the presence of dielectric
boundary forces in coarse-grained simulations cannot be neglected under many
conditions. Polarization effects due to the presence of dielectric mismatches at
interfaces can lead to important effects, as it has been demonstrated for the case of a
salt solution confined to a slit pore, as well as in the case of the translocation
free energy barrier of a model DNA through a nanopore. Although for the DNA-nanopore
system screening at finite salt concentration reduces the influence of the
dielectric boundary forces, in the case of a slit pore the effects cannot be
neglected even at relatively high (200 mmol/l) concentrations.

We gratefully acknowledge financial support by the Volkswagen
Foundation, by the
Deutsche Forschung Gemeindschaft through SFB 716 - TP C5, and by the
BMBF via the ScaFaCoS project. We thank the  
 \es\ team for support.

\bibliographystyle{model1a-num-names}
\bibliography{paper_ccp.bib}







\end{document}